\RequirePackage{fix-cm}
\documentclass[twocolumn,epjc3]{svjour3}  
\smartqed  
\RequirePackage{graphicx}
\RequirePackage{float}
\RequirePackage{hyperref}
\RequirePackage{amsmath}
\usepackage{widetext}
\RequirePackage{amssymb}
\RequirePackage{mathtools}      
\journalname{Eur. Phys. J. C}
\begin{document}

\title{Wormholes in exponential $f(R,T)$ gravity}


\author{P.H.R.S. Moraes\thanksref{e1,addr1,addr2}
        \and
        P.K. Sahoo\thanksref{e2,addr3}
             }

\thankstext{e1}{e-mail: moraes.phrs@gmail.com}
\thankstext{e2}{e-mail: pksahoo@hyderabad.bits-pilani.ac.in}


\institute{Universit\`a degli studi di Napoli ``Federico II'' - Dipartimento di Fisica, Napoli I-80126, Italy \label{addr1}
           \and
           ITA - Instituto Tecnol\'ogico de Aeron\'autica - Departamento de F\'isica, 12228-900, S\~ao Jos\'e dos Campos, S\~ao Paulo, Brasil \label{addr2}
          \and 
Department of Mathematics, Birla Institute of
Technology and Science-Pilani, Hyderabad Campus, Hyderabad-500078,
India \label{addr3}
}

\date{Received: date / Accepted: date}

\maketitle

\begin{abstract}
Alternative gravity is nowadays an extremely important tool to address some persistent observational issues, such as the dark sector of the universe. They can also be applied to stellar astrophysics, leading to outcomes one step ahead of those obtained through General Relativity. In the present article we test a novel $f(R,T)$ gravity model within the physics and geometry of wormholes. The $f(R,T)$ gravity is a reputed alternative gravity theory in which the Ricci scalar $R$ in the Einstein-Hilbert gravitational lagrangian is replaced by a general function of $R$ and $T$, namely $f(R,T)$, with $T$ representing the trace of the energy-momentum tensor. We propose, for the first time in the literature, an exponential form for the dependence of the theory on $T$. We derive the field equations as well as the non-continuity equation and solve those to wormhole metric and energy-momentum tensor. The importance of applying alternative gravity to wormholes is that through these theories it might be possible to obtain wormhole solutions satisfying the energy conditions, departing from General Relativity well-known outcomes. In this article, we indeed show that it is possible to obtain wormhole solutions satisfying the energy conditions in the exponential $f(R,T)$ gravity. Naturally, there is still a lot to do with this model, as cosmological, galactical and stellar astrophysics applications, and the reader is strongly encouraged to do so, but, anyhow, one can see the present outcomes as a good indicative for the theory. 

\end{abstract}

\section{Introduction}
\label{sec:int}

The $f(R,T)$ gravity theory was proposed in 2011 by T. Harko and collaborators \cite{harko/2011}. It starts from a generalization of the gravitational lagrangian of the $f(R)$ gravity \cite{sotiriou/2010,de_felice/2010}, in which besides the dependence on a general function of the Ricci scalar $R$, the gravitational lagrangian also depends generically on the trace of the energy-momentum tensor $T$.

The $f(R,T)$ theory has been applied to cosmology \cite{sharif/2018}-\cite{wu/2018} and stellar astrophysics \cite{sharif/2018b}-\cite{clmaomm/2017}, among other areas, yielding interesting and testable results.

The functional form of the function $f(R,T)$ is, in principle, arbitrary, though some attempts to constrain it have already been reported \cite{sharif/2013}-\cite{mirza/2016}.

In the present article we will present a new functional form for the function $f(R,T)$, namely $f(R,T)=R+\gamma e^{\chi T}$, with $\gamma$ and $\chi$ being constants. Such an exponential dependence on $T$ is a novelty in the $f(R,T)$ gravity literature. It is somehow motivated by exponential $f(R)$ gravity models \cite{odintsov/2017b}-\cite{campista/2011}.

We will test this form for the $f(R,T)$ function by solving the wormhole field equations in the formalism. 

Wormholes are shortcut tunnels that connect two different regions in space-time. They were proposed as a tool for teaching General Relativity \cite{morris/1988}, but now there is a number of possibilities to detect them \cite{shaikh/2018}-\cite{li/2014}.

The motivation for testing wormholes in extended gravity, such as the $f(R,T)$ theory, is that according to General Relativity, wormholes do not satisfy the energy conditions \cite{visser/1995}. Particularly, M.S. Morris and K.S. Thorne defined wormholes violating the weak energy condition as wormholes filled by ``exotic matter''. The extra degrees of freedom of extended gravity may allow the wormhole material solutions to satisfy one or more of the energy conditions \cite{mehdizadeh/2017}-\cite{sms/2018}.

In the present article we will obtain wormhole solutions in the $f(R,T)=R+\gamma e^{\chi T}$ gravity, with the purpose of checking if their material content is able to obey the energy conditions. In the next section, we will present the main features of the $f(R,T)$ gravity formalism and derive the field equations for the case $f(R,T)=R+\gamma e^{\chi T}$. In Section 3 we will substitute the wormhole metric and energy-momentum tensor in the $f(R,T)=R+\gamma e^{\chi T}$ gravity and derive the referred solutions. In Section 4 we present the energy conditions applied to the obtained material solutions and the discussions are presented in Section 5.

It is worth to clarify here that extended gravity theories have their main motivation based on solving the cosmological constant problem (check, for instance, \cite{flanagan/2001,zaripov/2014}) and predicting the accelerated expansion of the universe \cite{akrami/2013}-\cite{perivolaropoulos/2009}. Anyhow, extended gravity has also shown some remarkable results when applied to (extra-)galactic and stellar astrophysics, as it can be seen, respectively, in \cite{bohmer/2008}-\cite{capozziello/2013} and \cite{pani/2014}-\cite{novak/2000}. In a still more fundamental level, extended gravity theories may be the path to a quantum theory of gravity, as it has been profoundly discussed in Reference \cite{capozziello/2011}.

\section{The $f(R,T)$ gravity}

The $f(R,T)$ gravity starts from the total action \cite{harko/2011}

\begin{equation}\label{frt1}
S=\int d^4x\sqrt{-g}\left[\frac{1}{16\pi}f(R,T)+L\right].
\end{equation}
In \eqref{frt1}, $g$ is the determinant of the metric $g_{\mu\nu}$ and $L$ the matter lagrangian. Throughout the article, natural units will be assumed.

By varying the action above with respect to the metric yields the following field equations:

\begin{multline}\label{frt2}
f_R(R,T)R_{\mu\nu}-\frac{1}{2}f(R,T)g_{\mu\nu}+(g_{\mu\nu}\nabla^\mu\nabla_{\nu}-\nabla_\mu\nabla_\nu)f_R(R,T) \\=8\pi T_{\mu\nu}+f_T(R,T)(T_{\mu\nu}-Lg_{\mu\nu}).
\end{multline}
In \eqref{frt2}, $f_R(R,T)\equiv\partial f(R,T)/\partial R$, $R_{\mu\nu}$ is the Ricci tensor, $T_{\mu\nu}$ is the energy-momentum tensor and $f_T(R,T)\equiv\partial f(R,T)/\partial T$.

By applying the covariant derivative of \eqref{frt2} yields

\begin{multline}\label{frt3}
\nabla^{\mu}T_{\mu\nu}=\frac{f_T(R,T)}{8\pi+f_T(R,T)}\times \\ \left[(Lg_{\mu\nu}-T_{\mu\nu})\nabla^{\mu}\ln f_T(R,T)+\nabla^{\mu}\left(L-\frac{1}{2}T\right)g_{\mu\nu}\right].
\end{multline}

From \eqref{frt3} we observe that the energy-momentum tensor in principle does not conserve in $f(R,T)$ gravity. This may be related to quantum effects \cite{harko/2011}, such as particle creation \cite{harko/2014}, which is one of the motivations for inserting the $T$-dependence on the gravitational lagrangian. Anyhow, it is worth quoting that conserved versions of the $f(R,T)$ gravity have already been constructed \cite{mcr/2018,alvarenga/2013b,chakraborty/2013}.

\subsection{Exponential $f(R,T)$ gravity}

As mentioned early in the text, in the present article we are proposing a novel functional form for the $f(R,T)$ function. If it passes the wormhole energy conditions ``test'', the functional form is worth to be further applied to other areas, such as cosmological models, hydrostatic equilibrium of compact stars and rotation curves of galaxies, among others. The form is $f(R,T)=R+\gamma e^{\chi T}$. By substituting it in \eqref{frt2} yields 

\begin{equation}\label{efrt1}
G_{\mu\nu}=8\pi T_{\mu\nu}+\gamma e^{\chi T}\left[\frac{1}{2}g_{\mu\nu}+\chi(T_{\mu\nu}+pg_{\mu\nu})\right],
\end{equation}
as the field equations of the model, in which it has been assumed that $L=-p$, with $p$ being the total pressure. It is interesting to remark that the present functional form of $f(R,T)$ contains corrections only in the material sector of a gravity theory, so that the Einstein tensor is exactly the same as in General Relativity.

Also, by substituting the above exponential form for $f(R,T)$ in \eqref{frt3} yields

\begin{multline}\label{efrt2}
\nabla^{\mu}T_{\mu\nu}=\frac{-\gamma\chi e^{\chi T}}{8\pi+\gamma\chi e^{\chi T}} \times \\ \left[\chi(pg_{\mu\nu}+T_{\mu\nu})\nabla^\mu T+\nabla^\mu\left(p+\frac{1}{2}T\right)g_{\mu\nu}\right].
\end{multline}

\section{Wormhole equations in exponential $f(R,T)$ gravity}

The wormhole equations in exponential $f(R,T)$ gravity will be obtained from the substitution of the wormhole metric and energy-momentum tensor in Eqs.\eqref{efrt1}-\eqref{efrt2} above. The former reads \cite{morris/1988,visser/1995}

\begin{equation}\label{wh1}
ds^2=e^{2\Phi(r)}dt^2-\frac{1}{1-\frac{b(r)}{r}}dr^2-r^2(d\theta^2+\sin^2\theta d\phi^2),
\end{equation}
in which $\Phi(r)$ and $b(r)$ are, respectively, the redshift and shape functions. 

There are some conditions that must be respected by these functions in order for the wormhole to be traversable and asymptotically flat. Those are:

\begin{eqnarray}
\lim_{r\rightarrow\infty}\Phi(r)=\Phi_0,\label{wh2}\\
b(r_0)=r_0,\label{wh3}\\
b(r)<r,\label{wh4}\\
b'(r)<\frac{b(r)}{r}.\label{wh5}
\end{eqnarray}
In Eq.\eqref{wh2}, $\Phi_0$ is simply a finite number, in Eq.\eqref{wh3}, $r_0$ is the wormhole throat radius, \eqref{wh4} must be valid away from the throat and \eqref{wh5} near the throat. Throughout the article, primes indicate radial derivatives.

Let us calculate the Einstein tensor for the metric \eqref{wh1} above. When doing so we will consider constant $\Phi$, as it has been done in several references \cite{jamil/2010}-\cite{ms/2017}. A further study of wormholes in exponential $f(R,T)$ gravity may approach a non-constant redshift function. The non-null components of the Einstein tensor for metric \eqref{wh1} with constant $\Phi$ reads

\begin{eqnarray}
G_0^0=\frac{b'}{r^2},\label{wh6}\\
G_1^1=\frac{b}{r^3},\label{wh7}\\
G_2^2=\frac{b'r-b}{2r^3}.\label{wh8}
\end{eqnarray}

The material content of wormholes is described by an anisotropic matter energy-momentum tensor \cite{morris/1988,visser/1995}

\begin{equation}\label{wh9}
T_{\mu\nu}=\texttt{diag}(\rho,-p_r,-p_t,-p_t),
\end{equation}
in which $\rho$ is the matter-energy density and $p_r$ and $p_t$ are, respectively, the radial and transverse pressures.

By substituting \eqref{wh6}-\eqref{wh9} in \eqref{efrt1} yields

\begin{eqnarray}
\frac{b'}{r^2}=8\pi\rho+\gamma e^{\chi(\rho-p_r-2p_t)}\left[\frac{1}{2}+\chi\left(\rho+\frac{p_r+2p_t}{3}\right)\right],\label{wh10}\\
\frac{b}{r^3}=-4\pi p_r+\gamma e^{\chi(\rho-p_r-2p_t)}\left[\frac{1}{4}+\frac{\chi}{3}(p_t-p_r)\right],\label{wh11}\\
\frac{b'r-b}{2r^3}=-8\pi p_t+\gamma e^{\chi(\rho-p_r-2p_t)}\left[\frac{1}{2}+\frac{\chi}{3}(p_r-p_t)\right].\label{wh12}
\end{eqnarray}

Eq.\eqref{efrt2} for metric \eqref{wh1} and energy-momentum tensor \eqref{wh9} reads

\begin{multline}\label{wh13}
p_r'+\frac{2}{r}(p_r-p_t)=\frac{2\gamma\chi e^{\chi(\rho-p_r-2p_t)}}{3[8\pi+\gamma\chi e^{\chi(\rho-p_r-2p_t)}]} \times \\ \left[\chi(\rho'-p_r'-2p_t')(p_t-p_r)+\frac{1}{4}(3\rho'-p_r')-p_t'\right].
\end{multline}

\section{Wormhole equations of state, material solutions and energy conditions in exponential $f(R,T)$ gravity}

Let us consider the following relations 
\begin{eqnarray}
p_r=\alpha \rho, \label{wh14.1}\\
p_t=\beta \rho, \label{wh14.2}
\end{eqnarray}
as the equations of state for matter inside the wormholes, where $\alpha$ and $\beta$ are constants. Such relations have been invoked in other wormhole references, such as \cite{mehdizadeh/2017b},\cite{ms/2017}-\cite{garcia/2010}.

From Equations (\ref{wh11}) and (\ref{wh12}), we have

\begin{equation}\label{wh15}
\rho=\frac{\frac{\chi  (1-\zeta) (b+b' r)}{2 r^3}-4 \pi  \zeta W\left\{-\frac{3 \gamma  \chi  (1-\zeta) \exp \left[\frac{\chi  (1-\zeta) (b+b' r)}{8 \pi  r^3 \zeta}\right]}{16 \pi  \zeta}\right\}}{4 \pi  \chi  (1-\zeta) \zeta},
\end{equation}
where $\zeta\equiv\alpha+2\beta$ and $W$ denotes the Lambert function (also known as ``product logarithm''). 

Here, we will consider the power law form for the shape function as
\begin{equation}\label{wh16}
b=r_0^{m+1}r^{-m}.
\end{equation}
One can get different forms of shape function choosing different values for the parameter $m$. In \cite{lobo/2009}, some wormhole solutions in $f(R)$ gravity with the same shape function were investigated for $m=-1$ and $m=-\frac{1}{2}$. In order to satisfy the flaring out condition (10), we indeed need to consider $m<1$. In Fig.1 below we show the behaviour of $b(r)$ and $b'(r)$.

\begin{figure}[H]
\centering
\includegraphics[width=85mm]{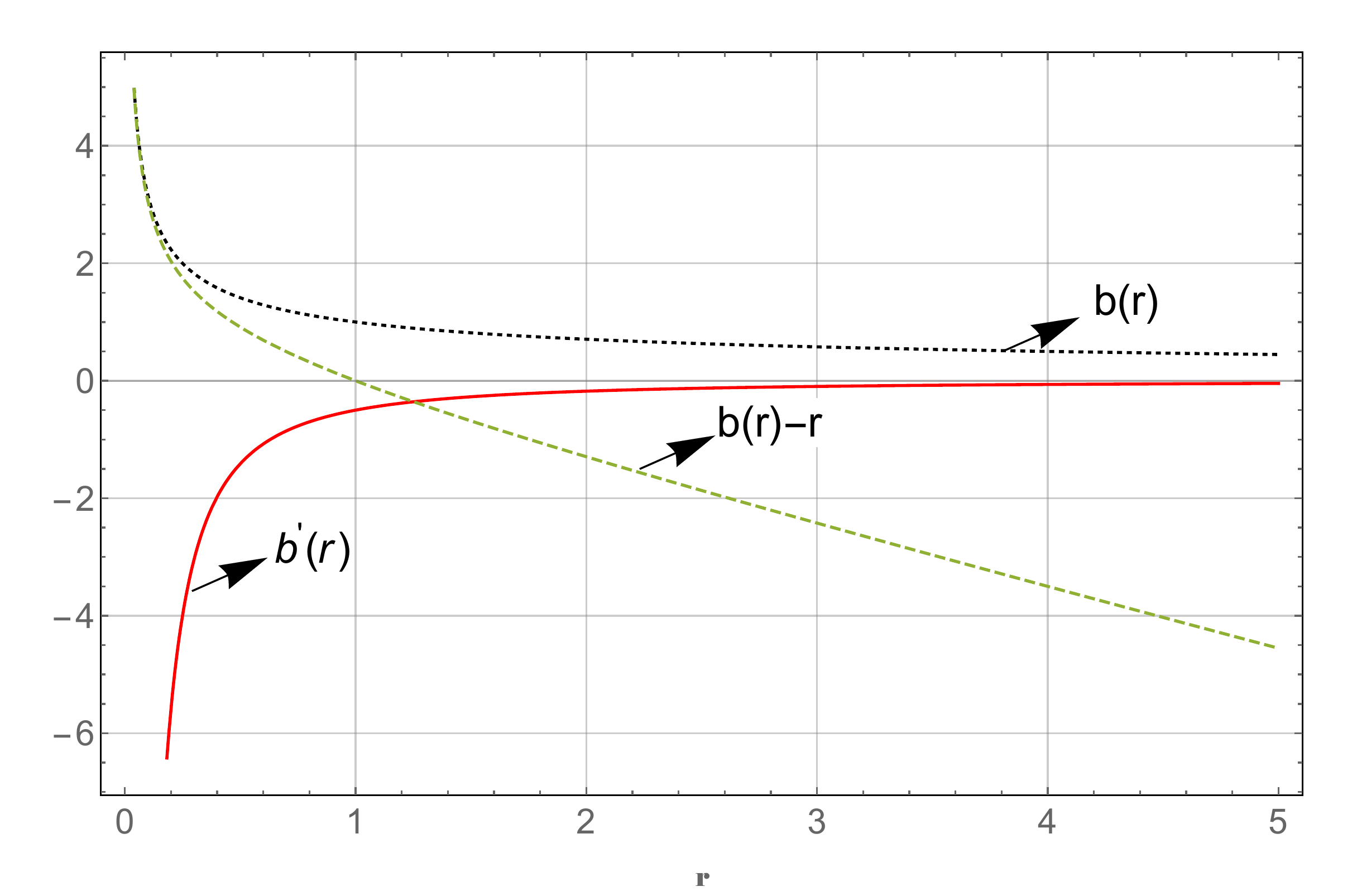}
\caption{Behaviour of the shape function $b(r)$ with $r_0=1$ and $m=0.5$.}\label{fig1}
\end{figure}

By using Equation (\ref{wh16}) in Equation (\ref{wh15}), we obtain obtain the following solution:

\newpage

\begin{widetext}
\begin{equation}\label{wh17}
\rho= \frac{\frac{\chi  (1-\zeta) r_0^{m+1} r^{-m}(1-m)}{2 r^3}-4 \pi \zeta W\left\{-\frac{3 \gamma  \chi  (1-\zeta) \exp \left[\frac{\chi  (1-\zeta) r_0^{m+1} r^{-m}(1-m)}{8 \pi  r^3 \zeta}\right]}{16 \pi  \zeta}\right\}}{4 \pi  \chi  (1-\zeta) \zeta}.
\end{equation}
\end{widetext}

Next we will analyse the behaviour of the material solutions as well as apply the energy conditions for different values of the $\alpha$ and $\beta$ parameters.

The energy-momentum tensor is not universal but depends on the particular type of matter and the interactions concerned. One of the key generic features that most matter we see experimentally seems to share is the positiveness of the energy density. 

The so-called energy conditions are a variety of ways of making the notion of locally positive energy density more precise. The energy conditions are very powerful techniques to deduce the theorems about the behaviour of strong gravitational fields and the geometries of cosmological models \cite{Visser/2000}. The point-like energy conditions take the form of various linear combinations of the components of the energy-momentum tensor at any specified point in space-time and those combinations should be positive or at least non-negative. Although they are not part of fundamental physics, they are useful to characterize the kind of fluid one deals with \cite{Barcelo/2002}. 

The energy conditions are \cite{Visser/1996}:

$\bullet$ Strong energy condition (SEC): the SEC says that gravity should always be attractive. In terms of the energy-momentum tensor components, it reads $ \rho +3p \geq 0 $.

$\bullet$ Dominant energy condition (DEC): the DEC indicates that the energy density measured by any observer should be non-negative, which leads to $\rho \geq \mid p \mid$.

$\bullet$ Weak energy condition (WEC): the WEC asserts that the energy density measured by any observer should be always non-negative, i.e., $\rho \geq 0$ and $\rho + p \geq 0$. 

$\bullet$ Null energy condition (NEC): the NEC is a minimum requirement from SEC and WEC, i.e. $\rho + p \geq 0$. The violation of NEC implies that all the above energy conditions are not validated. 

\subsection{Model A: equation of state with $\alpha=1/3$ and $\beta=1$}

\begin{figure}[H]
\centering
\includegraphics[width=80mm]{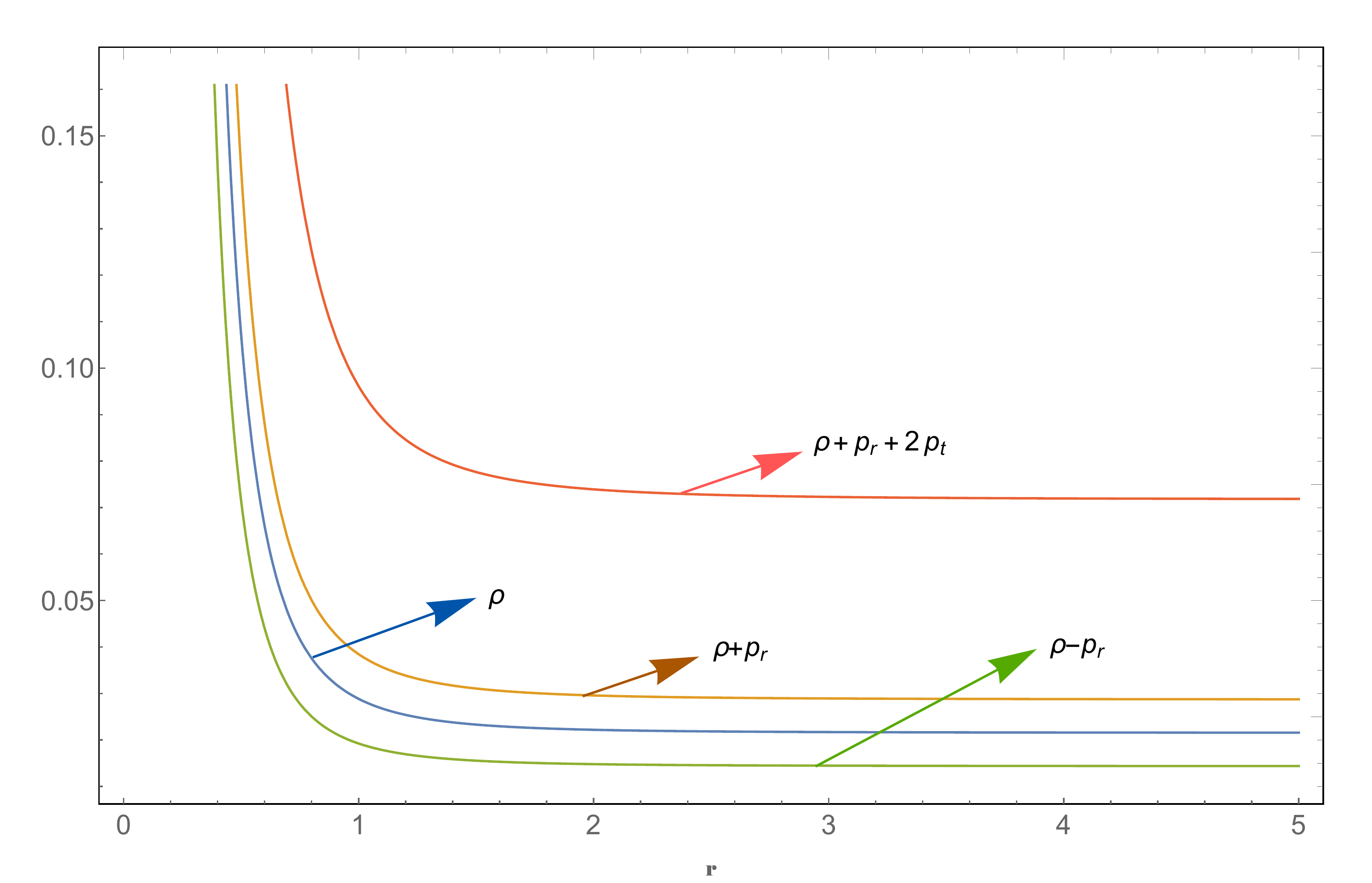}
\caption{Behaviour of the matter-energy density, NEC, DEC and SEC, with $r_0=1, \gamma=1$, $m=0.5$ and $\chi=6$, for model A.}\label{fig2}
\end{figure}

The matter-energy density of the wormhole constructed in the present formalism as well as its energy conditions in terms of $r$ are presented in Fig.\ref{fig2} below for Model A equation of state.

\subsection{Model B: equation of state with $\alpha=1$ and $\beta=1/2$}

The behaviour of the matter-energy density and the energy conditions in terms of $r$ in the present formalism wormhole for Model B equation of state are presented below, in Fig.\ref{fig3}.

\begin{figure}[H]
\centering
\includegraphics[width=80mm]{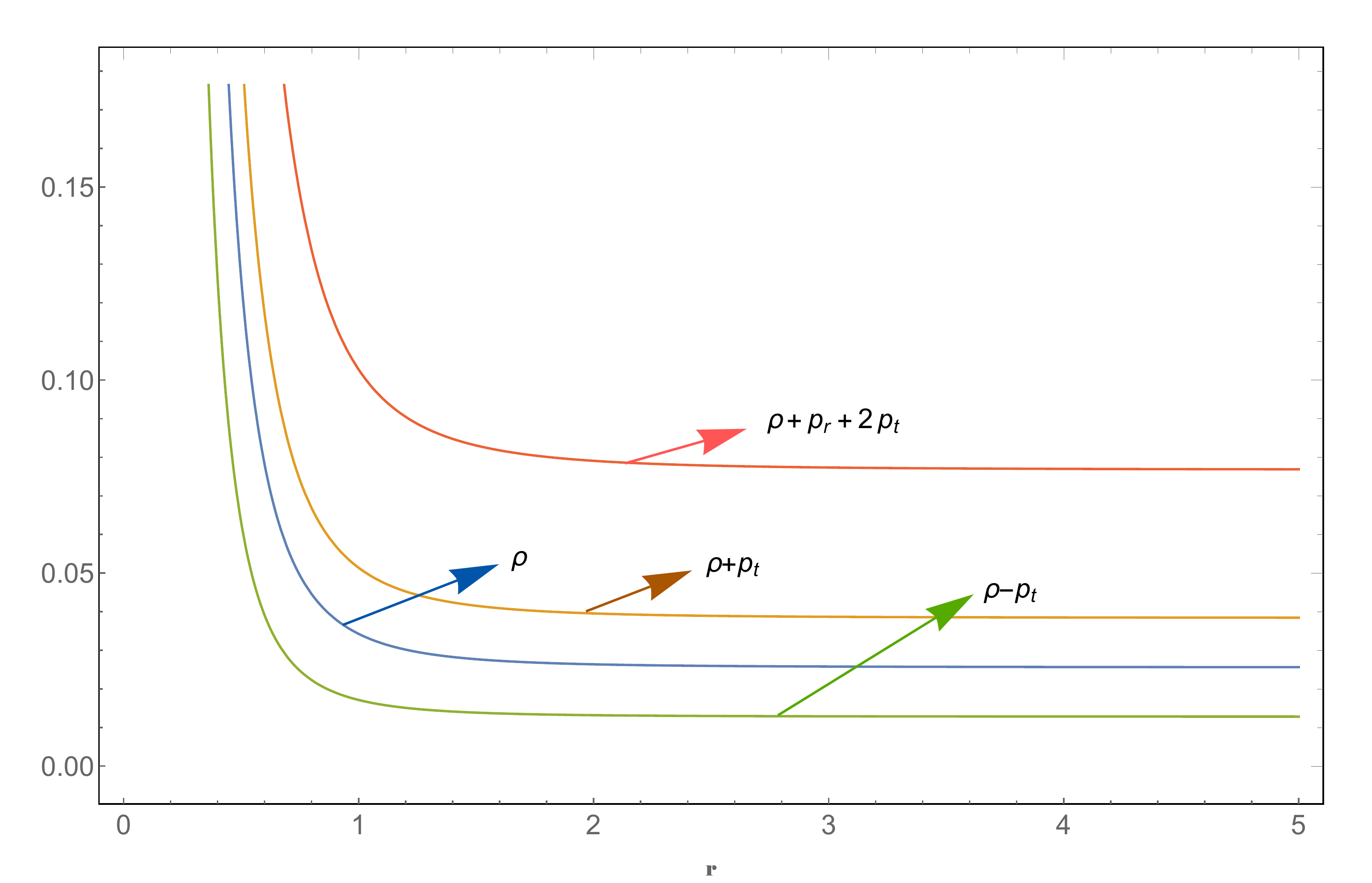}
\caption{Behaviour of the matter-energy density, NEC, DEC and SEC, with $r_0=1, \gamma=1$, $m=0.5$ and $\chi=6$, for model B.}\label{fig3}
\end{figure}

In the next section we discuss, among other features of the present model, the wormhole energy conditions presented in Figures \ref{fig2} and \ref{fig3}.

\section{Discussion}

The $f(R,T)$ theories of gravity have appeared firstly in 2011 \cite{harko/2011} and were proposed by Harko and collaborators as an alternative to the extended theories of gravity, since departing from most of them, it allows one to modify the material, rather than the geometrical sector, of General Relativity. One should note that when one assumes $f(R,T)=R+f(T)$, with $f(T)$ being a function of $T$ only, only the material sector of General Relativity is modified.

In fact, the $f(R,T)=R+f(T)$ models could well be mapped into models whose field equations read $G_{\mu\nu}=8\pi T_{\mu\nu}^{eff}$, with the effective energy-momentum tensor $T_{\mu\nu}^{eff}$ given by the energy-momentum tensor of a perfect fluid plus material corrections, such as bulk viscosity, anisotropy and other kinds of ``imperfections''. In other words, what these models do is to naturally insert correction terms to the material sector of a theory, by considering the possibility that the concept of perfect fluid may be an approximation. Particularly, this motivates the investigation of wormhole solutions within $f(R,T)$ models since wormholes contain imperfect anisotropic fluid filling them.

As a novelty in the literature, we have proposed in the present article an exponential dependence for $T$ in the $f(R,T)$ function, such as $f(R,T)=R+\gamma e^{\chi T}$. It is worth to recall that exponential $f(R)$ models are already present in the literature, as one can check \cite{odintsov/2017b}-\cite{campista/2011}. For the reasons we are going to present below, one can say that such a novel form has passed the wormhole test and we encourage further applications of it in other systems.

Although $f(R,T)$ models have been approved in some observational tests, an ultimate form for the function $f(R,T)$ still lacks. Nevertheless, the functional forms so far proposed for the $f(R,T)$ function are not able to predict some of the main observational issues today. Let us briefly analyse some present pros and cons of $f(R,T)$ models when compared to observations. 

It was shown in \cite{velten/2017} that only the conservative form of $f(R,T)$ gravity, namely $f(R,T)=R+\psi T^{1/2}$, with constant $\psi$, yields to a good matching between theory predictions and cosmological observations such as Type Ia Supernovae data. In \cite{clmaomm/2017} and \cite{mam/2016} it was shown that for the function $f(R,T)=R+2\lambda T$, with constant $\lambda$, the theoretically predicted maximum masses for white dwarfs and neutron stars indeed are increased but not sufficiently to be in touch with some observations \cite{howell/2006}-\cite{antoniadis/2013}. Furthermore, a deep and consistent analysis of galactic rotation curves for a particular $f(R,T)$ model still lacks in the literature.

As we mentioned above, the exponential $f(R,T)$ gravity has passed the wormhole test and below we explain the reasons why. The main motivation for working with wormholes within alternative gravity models is the possibility of obtaining wormhole solutions satisfying the energy conditions, departing from the General Relativity case \cite{morris/1988}. In fact, such a feature has already been attained through other alternative gravity theories \cite{mehdizadeh/2017}-\cite{mehdizadeh/2015}. 

Here, it was shown that through the exponential $f(R,T)$ gravity it is possible to obtain wormhole solutions remarkably  satisfying all the energy conditions and this can be seen in Figures \ref{fig2}-\ref{fig3}. Those figures were plotted for different cases of equations of state (\ref{wh14.1})-(\ref{wh14.2}). This shows that, in fact, the respectability of the energy conditions weakly depends on the equation of state and its parameters, but strongly depends on the background gravitational theory. 

Particularly, the importance of the background theory rather than the wormhole equation of state can also be seen for hybrid metric-Palatini gravity \cite{Capozziello/2012}.

It is important to stress here that a more profound discussion about energy conditions in alternative gravity was given in \cite{capozziello/2014,capozziello/2015}, in which the effective energy-momentum tensor of the theory is confronted with the energy conditions, rather than the usual one. In an extended gravity theory with extra material rather than geometrical terms, as the one presented here, in which the extra terms of the effective energy-momentum tensor also depend on $\rho$ and $p$, this approach is rather complicated to be attained, but we shall report it soon in the literature.

To finish, we mention that in Fig.1 we have also shown the shape function in terms of $r$ indicates that $b(r)<r$ is obeyed. The wormhole throat is located at $r_0=1$, which satisfies $b(r_0)=r_0$ as in Equation (\ref{wh3}). The radial derivative of the shape function $b'(r)$ satisfies Equations (\ref{wh4}) and (\ref{wh5}), i.e. $b'(r)<1$. Finally we also plotted $b(r)-r$ in Fig.1 and showed that it is $<0$ for $r>r_0$, also satisfying $1-b(r)/r>0$.

\textbf{Acknowledgments:} PHRSM would like to thank S\~ao Paulo Research Foundation (FAPESP), grants 2015/08476-0 and 2018/20689-7, for financial support. PKS acknowledges DST, New Delhi, India for providing facilities through DST-FIST lab, Department of Mathematics, where a part of this work was done. The authors also thank the referee for the valuable suggestions, which improved the presentation of the present results.

\end{document}